\documentclass[final,5p,times,twocolumn]{elsarticle}

\usepackage{graphicx}

\usepackage{amssymb}

\newcommand{\neqcm}{\ensuremath{\mathrm{n}_{\mathrm{eq}}/\mathrm{cm}^2}}
\newcommand{\mum}{\ensuremath{\mu}m}

\begin{document}

\begin{frontmatter}



\title{Development of active edge pixel sensors and four-side buttable modules using vertical integration technologies }


\author[A]{A. Macchiolo}
\author[B]{L. Andricek} 
\author[A]{H.-G Moser}
\author[A]{R. Nisius}
\author[B]{R.H. Richter}
\author[A]{S. Terzo}
\author[A]{P. Weigell}

\address[A]{Max-Planck-Institut for Physics, F\"{o}hringer Ring 6, D-80805 Munich, Germany}
\address[B]{Semiconductor Laboratory of the Max-Planck-Society, Otto Hahn Ring 6, D-81739 Munich, Germany}

\begin{abstract}

We present an R\&D activity focused on the development of novel modules for the upgrade of the ATLAS pixel system at the High Luminosity LHC (HL-LHC).  The modules consist of n-in-p pixel sensors, 100 or 200 $\mu$m thick, produced at VTT (Finland) with an active edge technology, which considerably reduces the dead area at the periphery of the device. The sensors are interconnected with solder bump-bonding to the ATLAS FE-I3 and FE-I4 read-out chips, and characterized with radioactive sources and beam tests at the CERN-SPS and DESY. The results of these measurements will be discussed for devices before and after irradiation up to a fluence of  $5\times 10^{15}$ \neqcm.
We will also report on the R\&D activity to obtain Inter Chip Vias (ICVs) on the ATLAS read-out chip in collaboration with the Fraunhofer Institute EMFT.  This step is meant to prove the feasibility of the signal transport to the newly created readout pads on the backside of the chips allowing for four side buttable devices without the presently used cantilever for wire bonding. The read-out chips with ICVs will be interconnected to thin pixel sensors, 75 $\mu$m and 150 $\mu$m thick, with the Solid Liquid Interdiffusion (SLID) technology, which is an alternative to the standard solder bump-bonding.

\end{abstract}


\end{frontmatter}


\section{Introduction}

The ATLAS pixel system will undergo around the years 2022-2023 a complete replacement to cope with the higher detector occupancy and
radiation doses  foreseen in the High Luminosity phase of the LHC (HL-LHC) \cite{HL-LHC}.
The possibility of employing thin pixel detectors is very attractive for the inner layers 
of the upgraded ATLAS pixel system, given the reduced material usage that helps minimizing
the multiple scattering experienced by charged particles. Other advantages offered by thin pixels are 
the good charge collection efficiency after irradiation, thanks to the higher electric field
that can be established in the silicon bulk with respect to thicker devices,
and the possibility of reducing the number of clusters with more than
two hit pixels, that are detrimental for the position resolution and the 
detector occupancy. This is particularly important
for the inner layers, where more  particles traverse the pixel modules at  high pseudo-rapidity ($\eta$). 
For example, at the maximum $\eta$ value of 2.5 covered by the ATLAS Insertable B-Layer \cite{IBL},
250 $\mu$m thick sensors, with 250 $\mu$m pitch along the beam direction, yield a mean cluster 
size of 7.1, while 150 $\mu$m thick sensors a mean cluster size of 4.6, as reported in \cite{IWORID}.

The performance of 75 and 150 $\mu$m thin n-in-p pixel sensors produced at
the Semiconductor Laboratory of the Max-Planck Society are discussed in \cite{Pixel2012}.
 In particular, it has been shown, that a charge collection efficiency (CCE) of (90$\pm$9)\% was obtained at a bias voltage of 750 V
after irradiation at a fluence of $10^{16}$ \neqcm\ for the 75 $\mu$m thin sensors 
interconnected to FE-I3 chips \cite{FE-I3}. 
In addition, it has also been reported in \cite{Pixel2012} that the 150 $\mu$m thin sensors, interconnected to FE-I4 chips \cite{FE-I4}, after being irradiated
at a fluence of $4\times 10^{15}$ \neqcm, yield hit efficiencies
of (96.5$\pm$0.3)\%  at 400V and (96.9$\pm$0.3)\% at 500V, for perpendicular beam incidence.

\section{Active edge pixels}
An additional requirement to instrument the inner pixel layers is the reduction of the module
inactive area due to space constraints that do not allow for overlapping the sensors along
the beam direction. Different approaches have been followed to achieve this goal 
for planar pixel detectors, as documented in \cite{PPS_philipp}.
We report here on the results obtained with Deep Reactive Ion Etching (DRIE), 
to achieve trenches around the sensors, which allow for
a doping of the sensor sides. This technology has been used for a multi project production of 
active edge n-in-p pixel sensors at VTT on 6" wafers. 
These devices were bump-bonded at VTT to FE-I3 and FE-I4 chips, and the results of the first electrical 
characterization have been reported in \cite{Pixel2012}. 
The bulk material, where not explicitely otherwise indicated, is Float Zone (FZ) p-type silicon,
with an initial resistivity of 10 k$\Omega$ cm.
The fabrication of thin sensors at VTT exploits
the use of a handle wafer as mechanical support during the
grinding phase. Trenches are created at the sensor border by wet etching, and then
activated by means of four-quadrant ion implantations with boron \cite{VTT1,VTT2}.
In the pixel production discussed in this paper, sensors were manufactered with a thickness
of 100 or 200 $\mu$m. At the end of the sensor production the handle wafer is removed
before the interconnection via solder bump-bonding to the chips.
An homogeneous p-spray method, with an uniform low-dose boron implantation,
has been used to to achieve the inter-pixel isolation.

\begin{figure}[ht]
\centering
\includegraphics[width=\columnwidth]{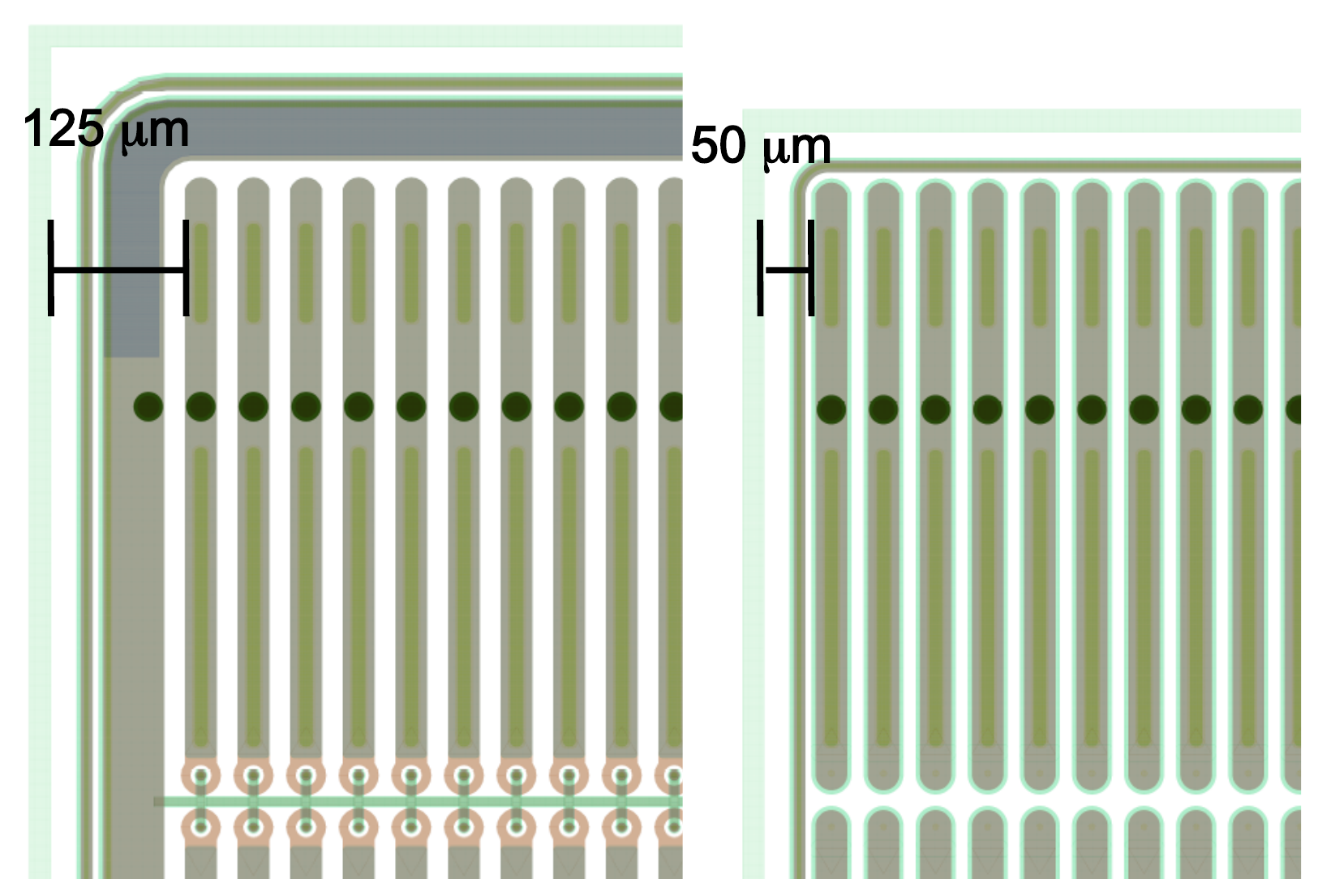}
\caption{Edge designs for VTT sensors: on the left picture, one bias ring 
and one guard ring with d$_\mathrm{edge}$=125 \mum, 
implemented in FE-I3 and FE-I4 modules; on the right picture, only one guard ring with d$_\mathrm{edge}$=50 \mum,
implemented in FE-I3 modules.}
\label{fig:VTTedges}
\end{figure}

Two versions of reduced edge design have been implemented,
as shown in Fig.\ref{fig:VTTedges}.
The first one is characterized by one guard and one bias ring,
with the latter connected to the pixel punch-through structures, allowing
for testability before interconnection. In this case the edge width, d$_\mathrm{edge}$, defined as the distance
between the end of the last pixel implant and the trench, is equal to 125 $\mu$m.
The second, more aggressive design, implemented only for FE-I3 sensors, foresees
one floating guard-ring and d$_\mathrm{edge}$=50 $\mu$m. 

\begin{figure}[hbt] 
\centering 
\includegraphics[width=\columnwidth,keepaspectratio]{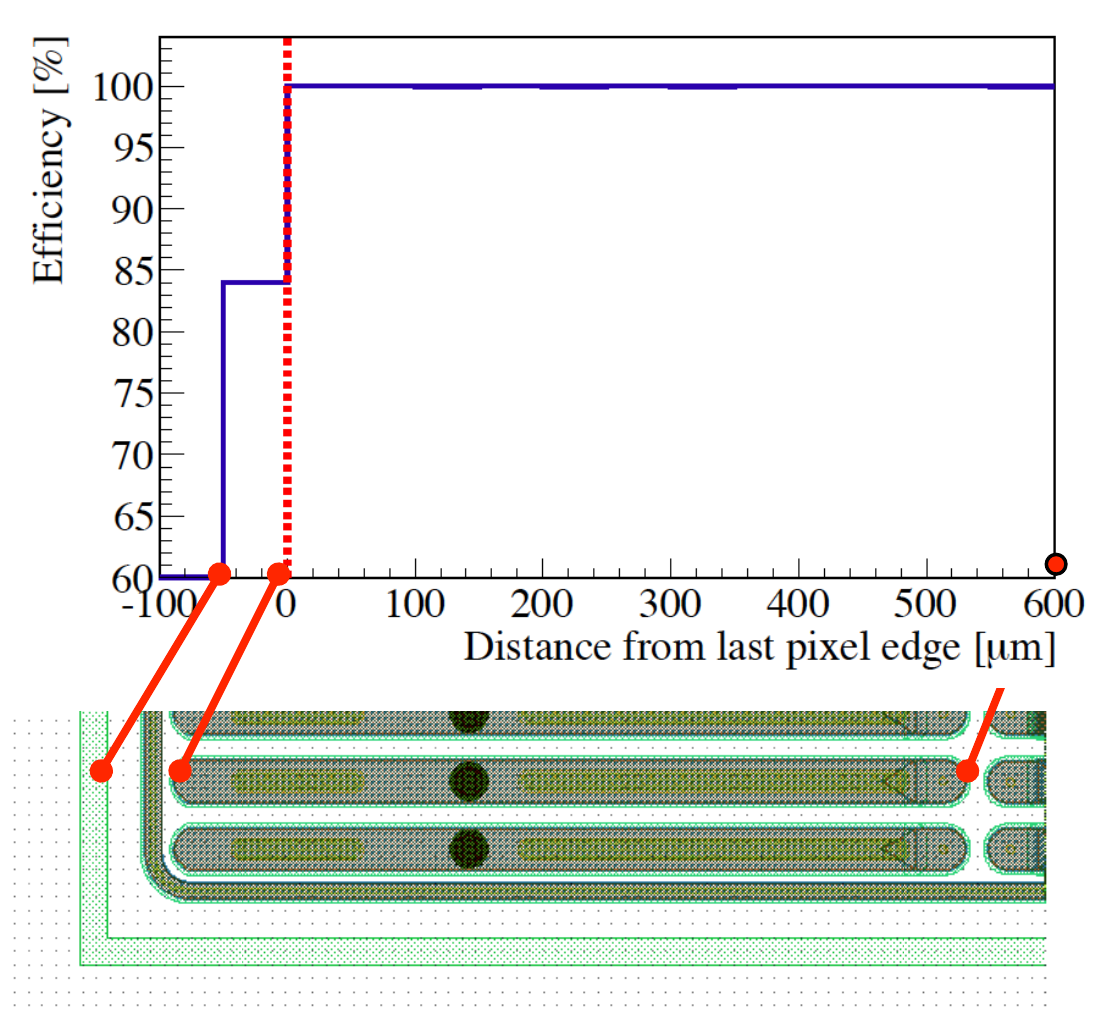}
\caption{Hit efficiency for the tracks crossing the edge pixel column, 
as a function of the distance from the end of the pixel implant. In this version of the FE-I3 sensor,
 the pixels in the last column are 600 $\mu$m long.} 
\label{fig:EdgeEfficiency}
\end{figure}

The absence of the bias ring also leads to the omission of the bias dot and bias
rail structures. These devices, before irradiation, have been characterized in a beam test at the CERN-SPS
with 120 GeV pions.
The global module efficiency and the hit efficiency at the sensor edge have been extracted.
The FE-I3 and FE-I4 modules yield a global efficiency of $(99.9^{+0.1}_{-0.3})$\% and $(99.8^{+0.2}_{-0.3})$\% respectively, at a bias
voltage of 20 V. The dominant source of the uncertainty for these measurements is systematic, and it has been evaluated 
as described in \cite{test-beam}.
Fig.\ref{fig:EdgeEfficiency} shows the hit efficiency
obtained over the edge pixel column, as a function of the distance from  
the end point of the pixel implant, for the sensor design with only a guard-ring.
A value of $(99.9^{+0.1}_{-0.3})$\% is found over the entire lenght  
of the pixel implantation, while in the 50 $\mu$m wide edge, a hit efficiency of (84$^{+9}_{-14}$)\%,
where the uncertainty is statistically dominated,
demonstrates that the sensor is still active also in this region. 
A detailed analysis of the edge efficiency for different geometries can be found in \cite{IWORID}.

\begin{figure}[hbt] 
\centering 
\includegraphics[width=\columnwidth,keepaspectratio]{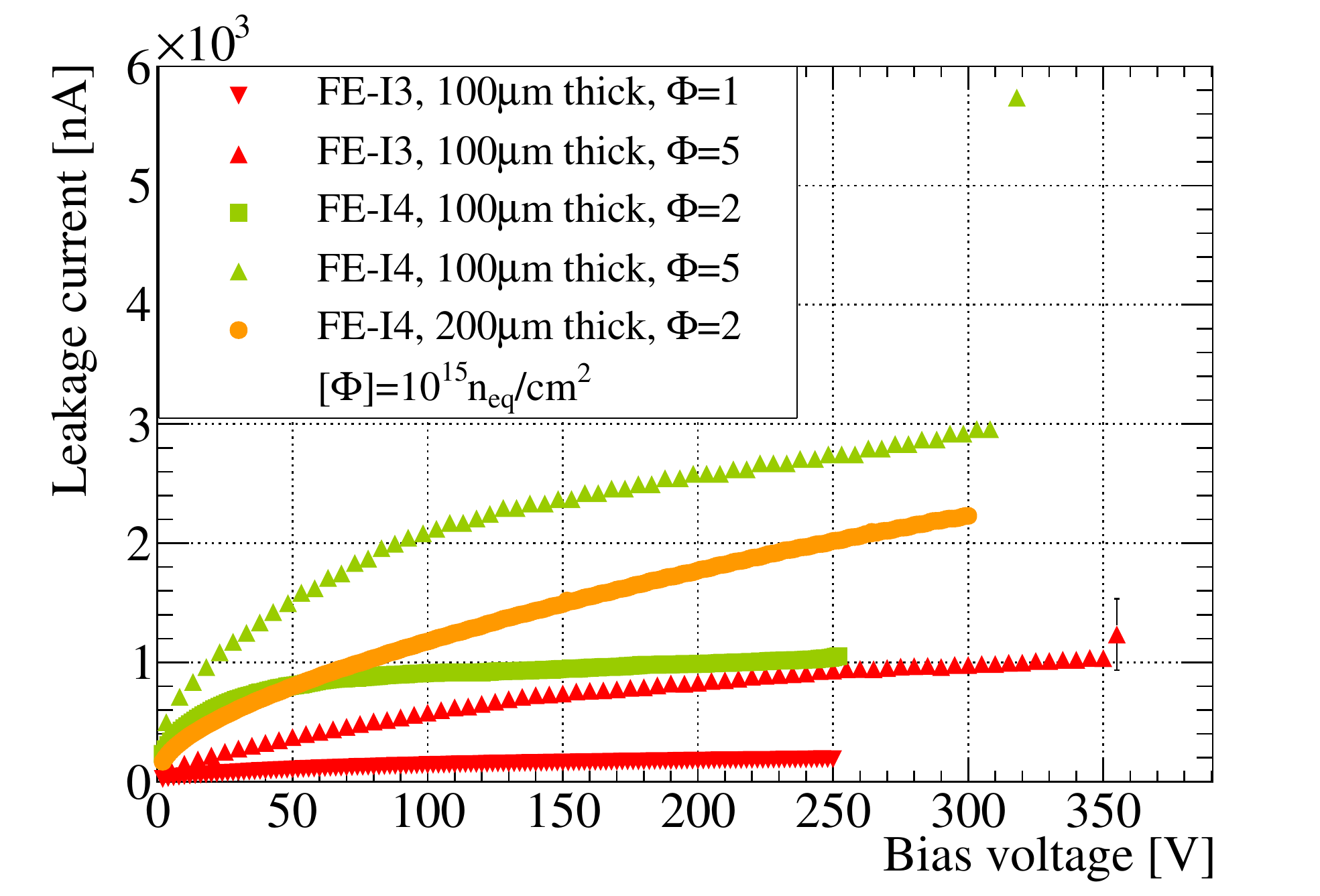}
\caption{IV curves of the VTT pixel modules with active edges after irradiation. The measurements were performed
inside a climate chamber at an environmental temperature of T=-50 $^{\circ}$C .}
\label{fig:IV_afterirr}
\end{figure}

\begin{figure}[hbt] 
\centering 
\includegraphics[width=\columnwidth,keepaspectratio]{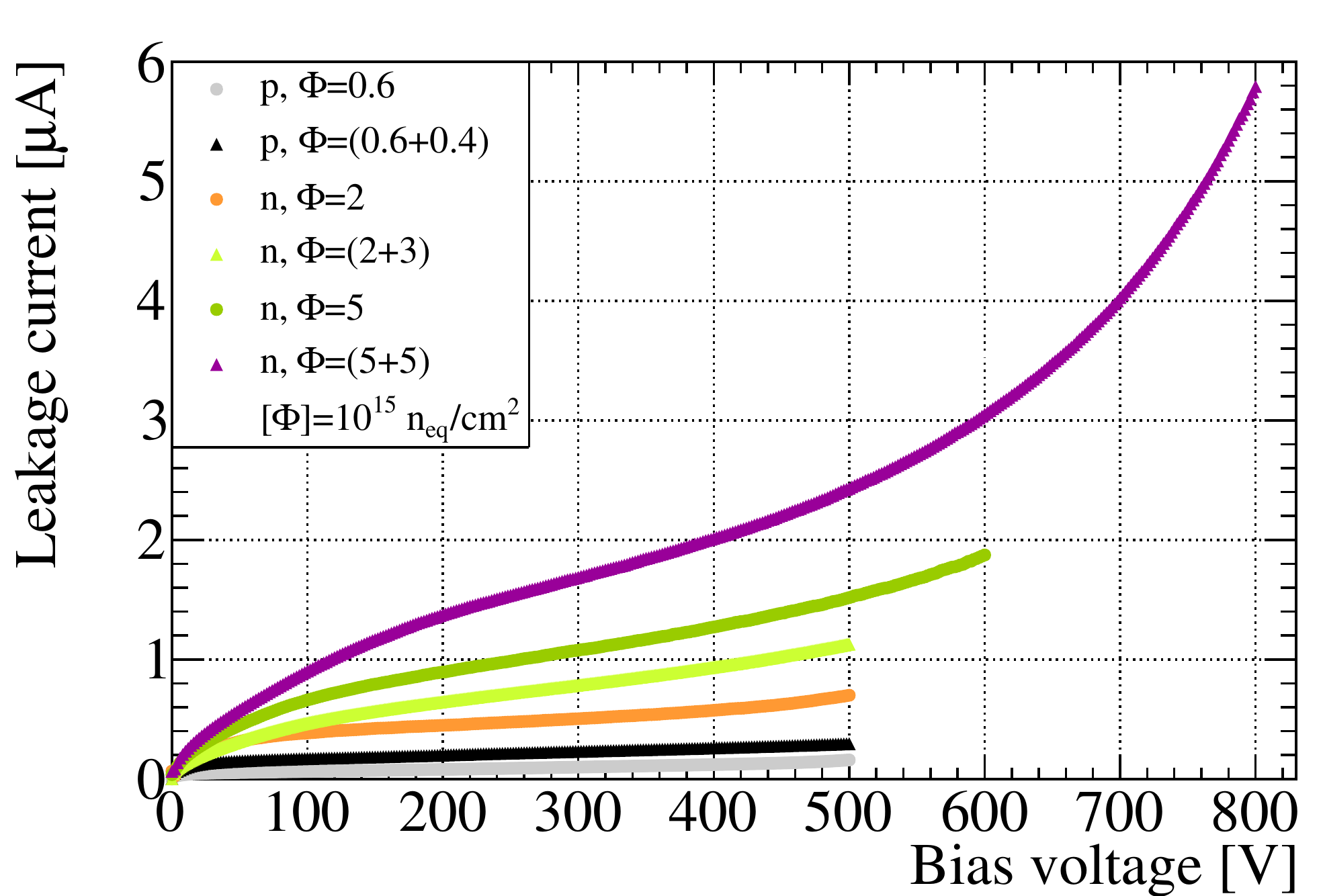}
\caption{IV curves of pixel modules built with 75 $\mu$m thin sensors interconnected to FE-I3 chips, irradiated
with reactor neutrons at JSI, from \cite{thesis_philipp}}
\label{fig:SLID_IV}
\end{figure}

\begin{figure}[h] 
\centering 
\includegraphics[width=\columnwidth,keepaspectratio]{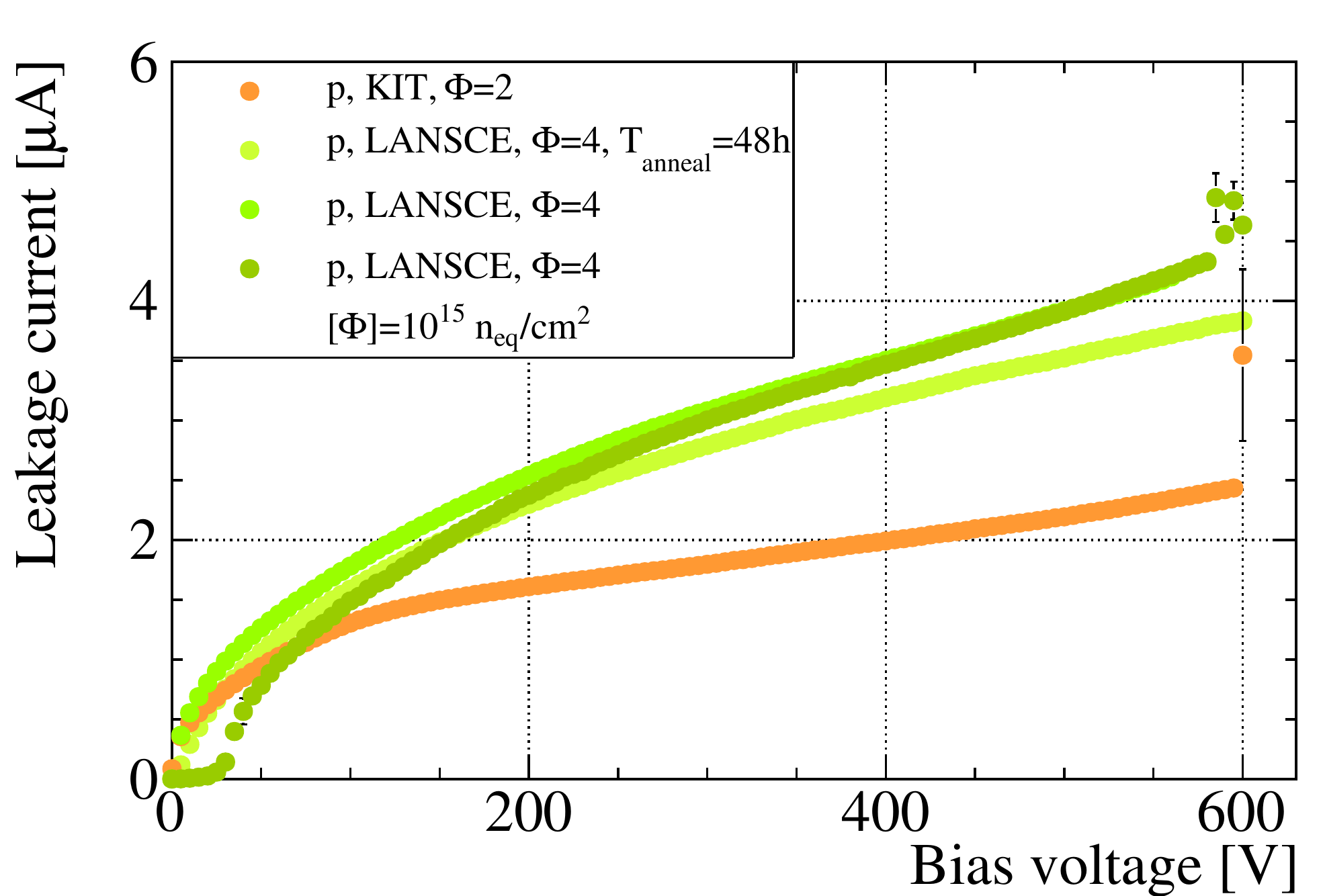}
\caption{IV curves of pixel modules built with 150 $\mu$m thin sensors, interconnected to FE-I4 chips
and irradiated with 25 MeV protons at KIT and 800 MeV protons at Los Alamos National Laboratory (LANSCE), 
from \cite{thesis_philipp}. T$_\mathrm{anneal}$ indicates the additional annealing time of one sample with
respect to the others. }
\label{fig:HLL_IV}
\end{figure}

Some samples of the active edge pixel devices  were irradiated 
at different fluences. A summary of the leakage currents measured for these sensors is presented in 
Fig.\ref{fig:IV_afterirr}. As expected, the current values scale according to the thickness and 
the fluence received. The breakdown voltages, indicated by the rightmost data point, 
 are larger than before irradiation \cite{Pixel2012}, with the highest value
of 350 V corresponding to the sample irradiated to the highest fluence achieved so far of $5\times 10^{15}$ \neqcm.
However, the breakdown voltages for these samples are generally lower than those measured
for thin n-in-p pixels devices of other productions with a guard-ring structure,  irradiated at the same fluences.
For comparison the IV curves of 75 $\mu$m thin detectors with 1 mm wide guard ring structures and of 
150 $\mu$m detectors, with a 450 $\mu$m wide structures, are reported in Fig.\ref{fig:SLID_IV} and 
Fig.\ref{fig:HLL_IV} \cite{thesis_philipp}.  

\begin{figure}[hbt] 
\centering 
\includegraphics[width=\columnwidth,keepaspectratio]{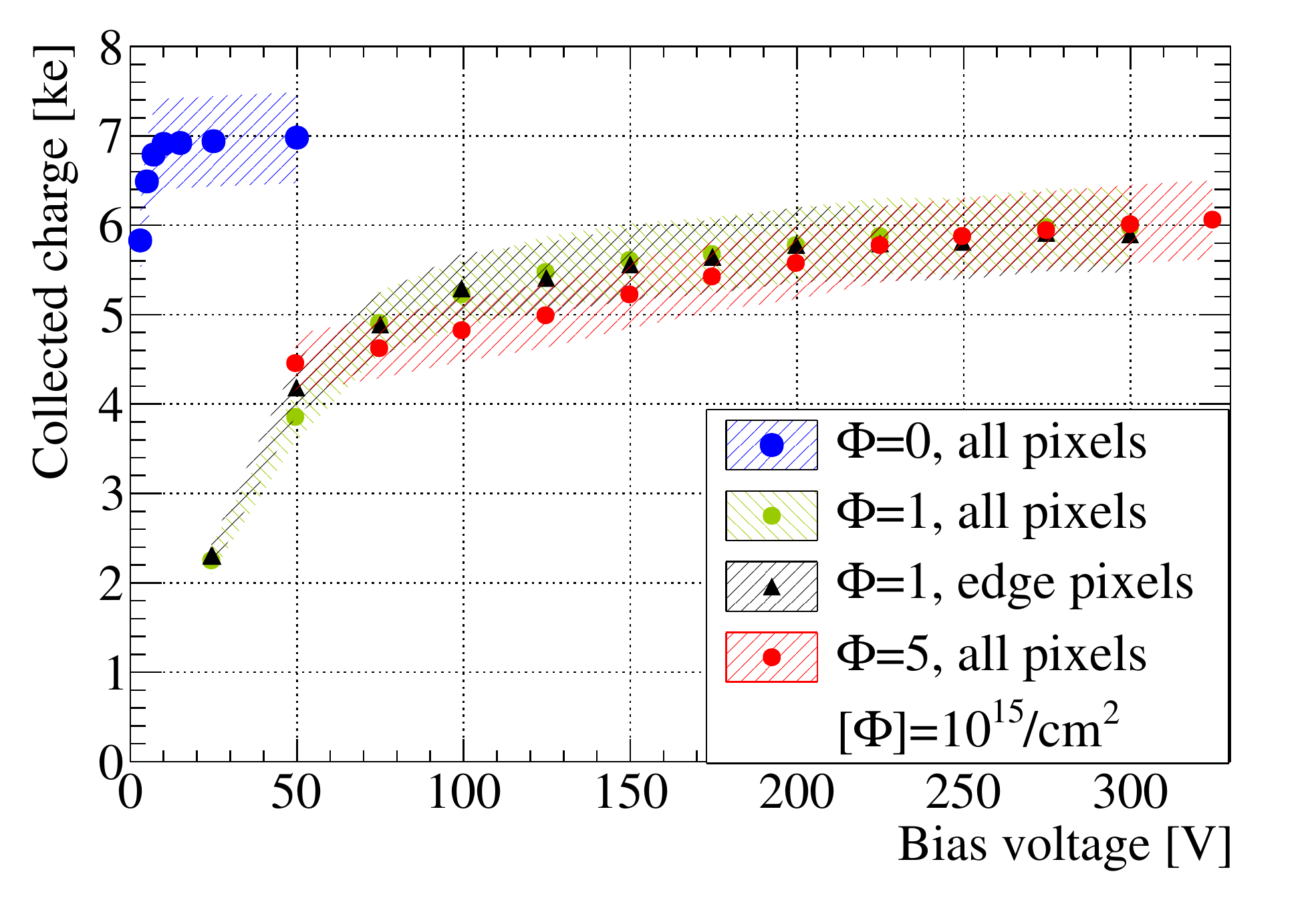}
\caption{Collected charge of a VTT FE-I3 module, with an edge width of 125 $\mu$m, irradiated first at KIT with 25 MeV protons 
at $\Phi$=$10^{15}$ \neqcm\ and then  with reactor neutrons at JSI at $\Phi$
of $4 \times 10^{15}$ \neqcm.
The blue curve corresponds to the charge collected before irradiation.
The green  and black curves correspond to the charge collected by all the pixels and only the edge pixels in the device at $\Phi$=$10^{15}$ \neqcm\ while the red curve corresponds to all the pixels at $\Phi$=$5 \times 10^{15}$ \neqcm}.
\label{fig:FEI3KIT}
\end{figure}

All the measurements of charge collection described in this paper were performed 
by using the  ATLAS USBPix read-out system \cite{USBPix}.  The modules
after irradiation were tested at an environmental temperature of T=-50$^{\circ}$C inside a climate
chamber to reproduce the conditions during the test-beam where the cooling is realized 
with dry ice. The Time over Threshold (ToT) to charge calibration was corrected by using measurements with
$^{109}$Cd and $^{241}$Am radioactive sources. A  residual systematic error of 10\% 
was assigned to the charge values extracted in runs with a $^{90}$Sr source.
A VTT FE-I3 pixel module with an edge width of 125 $\mu$m has been irradiated at the
Karlsruhe Institute of Technology (KIT) with 25 MeV protons 
at a fluence of $10^{15}$ \neqcm,
and tested with a $^{90}$Sr source up to a bias voltage of 300 V.
Fig.\ref{fig:FEI3KIT} shows that the collected charge in the edge pixels (black curve) is comparable to 
to the charge collected in the inner part of the module (green curve), similarly to what was observed before irradiation \cite{Pixel2012}.
At V$_\mathrm{bias}$=300 V, the collected charge is (86$\pm$9)\% of the value collected
before irradiation. The same module was afterwards irradiated with reactor neutrons at the
Jo\v{z}ef Stefan Institute (JSI) in Slovenia at a fluence of 
$4 \times 10^{15}$ \neqcm, for a total received fluence of $5 \times 10^{15}$ \neqcm. 
The charge collection properties at this higher irradiation fluence did not degrade further and the corresponding results are 
reported as the red curve in Fig. \ref{fig:FEI3KIT}. 

\begin{figure}[hbt] 
\centering 
\includegraphics[width=\columnwidth,keepaspectratio]{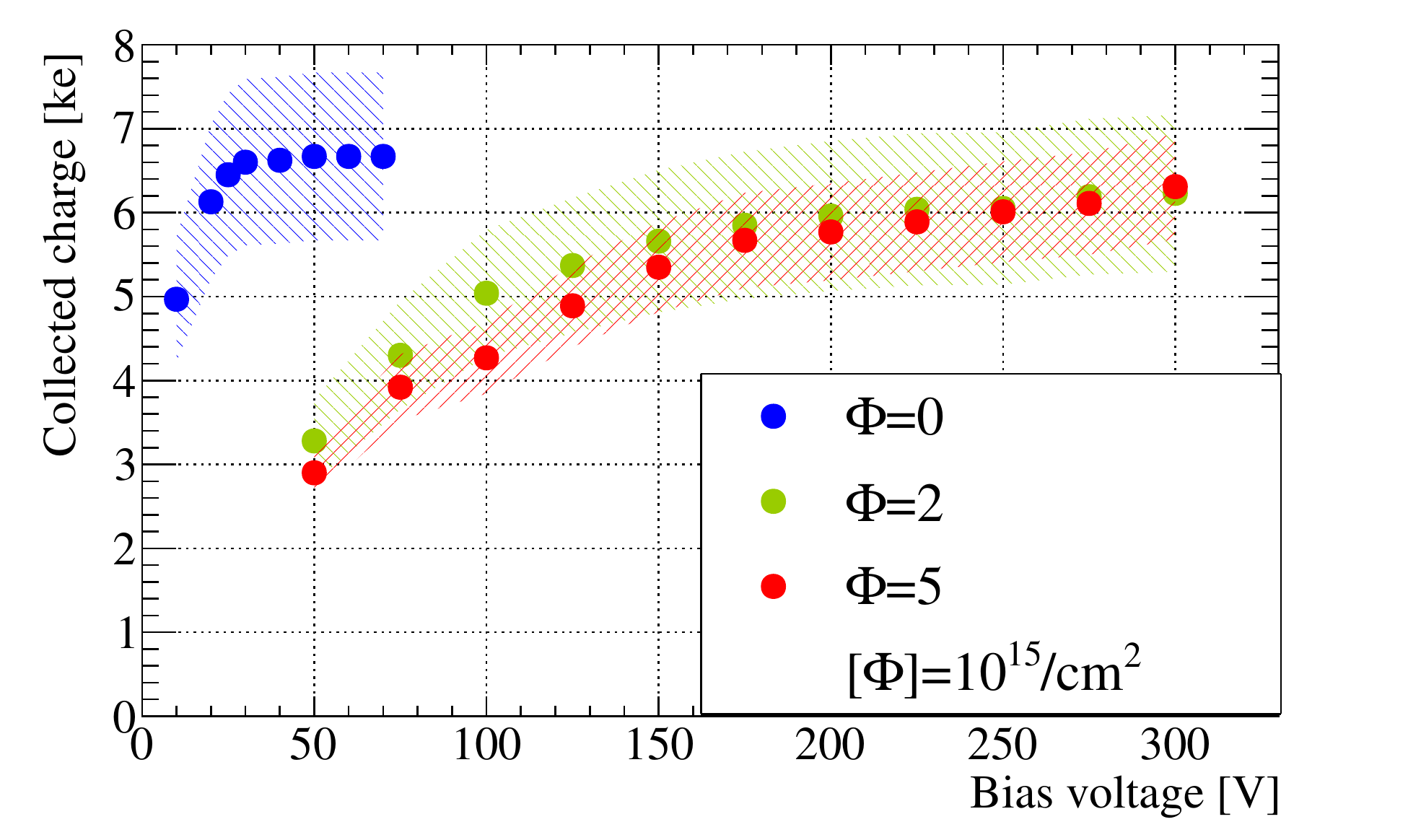}
\caption{Collected charge of a VTT FE-I4 module with a MCz silicon sensor, 100 $\mu$m thick, irradiated first at KIT with 25 MeV protons 
at $\Phi$=$2 \times 10^{15}$ \neqcm\ and then at $\Phi$=$5 \times 10^{15}$ \neqcm.
The blue curve corresponds to the charge collected before irradiation.}
\label{fig:MCZ_CCE}
\end{figure}

\begin{figure}[t] 
\centering 
\includegraphics[width=\columnwidth,keepaspectratio]{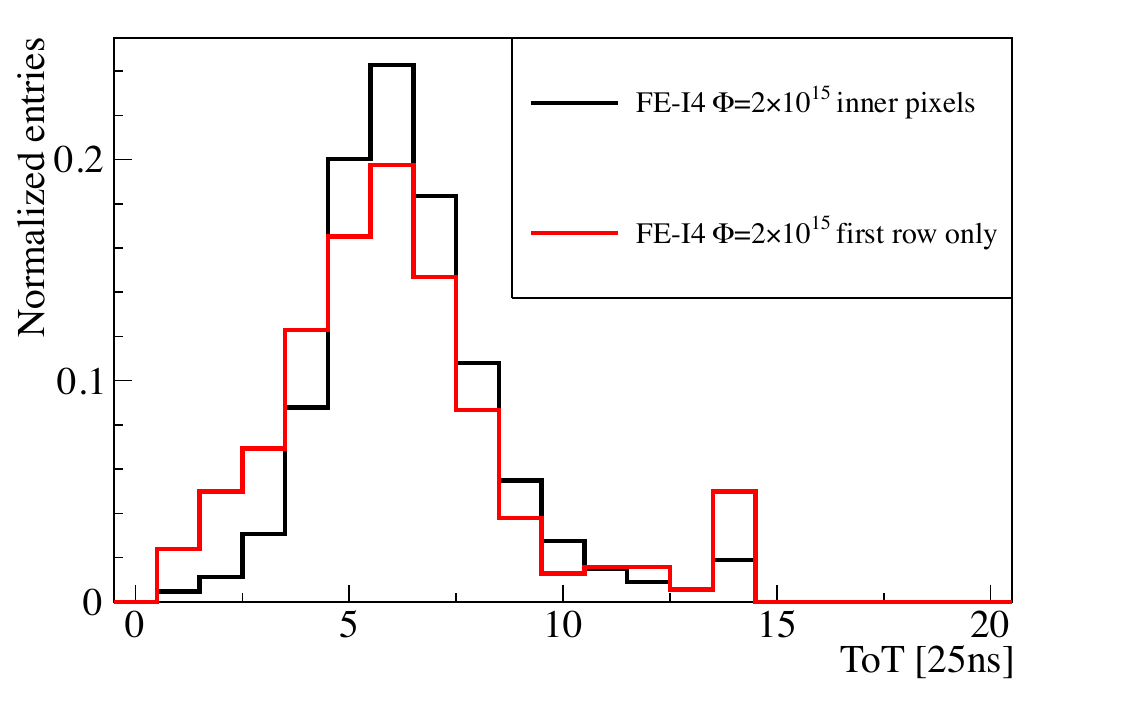}
\caption{Landau distribution of the collected charge, expressed in ToT,  obtained with a FE-I4 module with a 150 $\mu$m wide edge, irradiated at a fluence
of $2 \times 10^{15}$ \neqcm. The charge collected by the first row, in red, is similar  to the charge collected by the inner
pixels, in black. The slightly lower value can be explained by the loss of charge in the neighbouring bias ring, at ground potential,
that is not read-out.}
\label{fig:MCZ_Landau}
\end{figure}

A confirmation of the good charge collection properties of these thin detectors was achieved with 
a FE-I4 sensor, also 100~$\mu$m thick and with a 125 $\mu$m wide edge. In this case the bulk material
consists of MCz silicon, with orientation $\langle$100 $\rangle$ and an initial resistivity of 2 K$\Omega$ cm.
The module was irradiated at KIT with 25 MeV protons  in two successive steps first
at a fluence of $2 \times 10^{15}$ \neqcm\ and then at a total fluence of
$5 \times 10^{15}$ \neqcm.
The charge collection with a $^{90}$Sr source is compared before and after irradiation in Fig.\ref{fig:MCZ_CCE}.
At the fluence of $5\times 10^{15}$ \neqcm, with 300V of bias voltage, (91$\pm$9)\% of the pre-irradiation charge 
is achieved.
The Landau distributions of the collected charge for the internal and the edge pixels,
at a bias voltage of 300 V and at a fluence of $2 \times 10^{15}$ \neqcm,
are presented in Fig.\ref{fig:MCZ_Landau}.  The measurement was performed using a $^{90}$Sr source. Also for this device the edge pixels show a 
comparable charge collection performance  with respect to the internal ones.

\begin{figure}[h] 
\centering 
\includegraphics[width=\columnwidth,keepaspectratio]{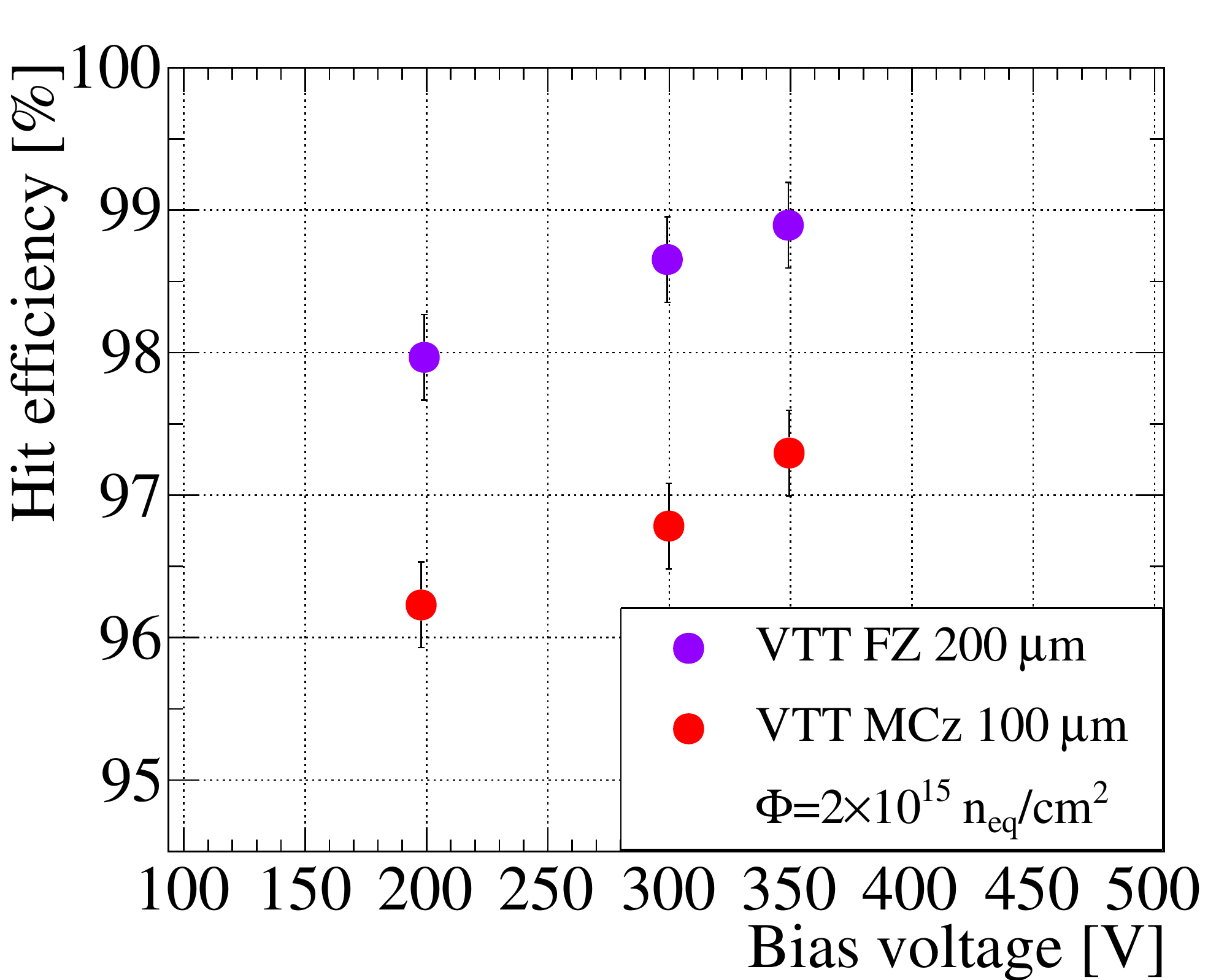}
\caption{Hit efficiency, as a function of the applied bias voltage, of two pixel modules, with a sensor thickness of 100 and 200 $\mu$m, irradiated at a fluence of $\Phi$=$2 \times 10^{15}$ \neqcm.}
\label{fig:irreff}
\end{figure}

These irradiated detectors were tested with a 4 GeV electron beam at DESY, using the EUDET telescope
for track reconstruction \cite{test-beam}.  Hit efficiencies, as a function of the bias voltage, were derived for the pixel modules, as shown in Fig.\ref{fig:irreff}.

\begin{figure}[h] 
\centering 
\includegraphics[width=\columnwidth,keepaspectratio]{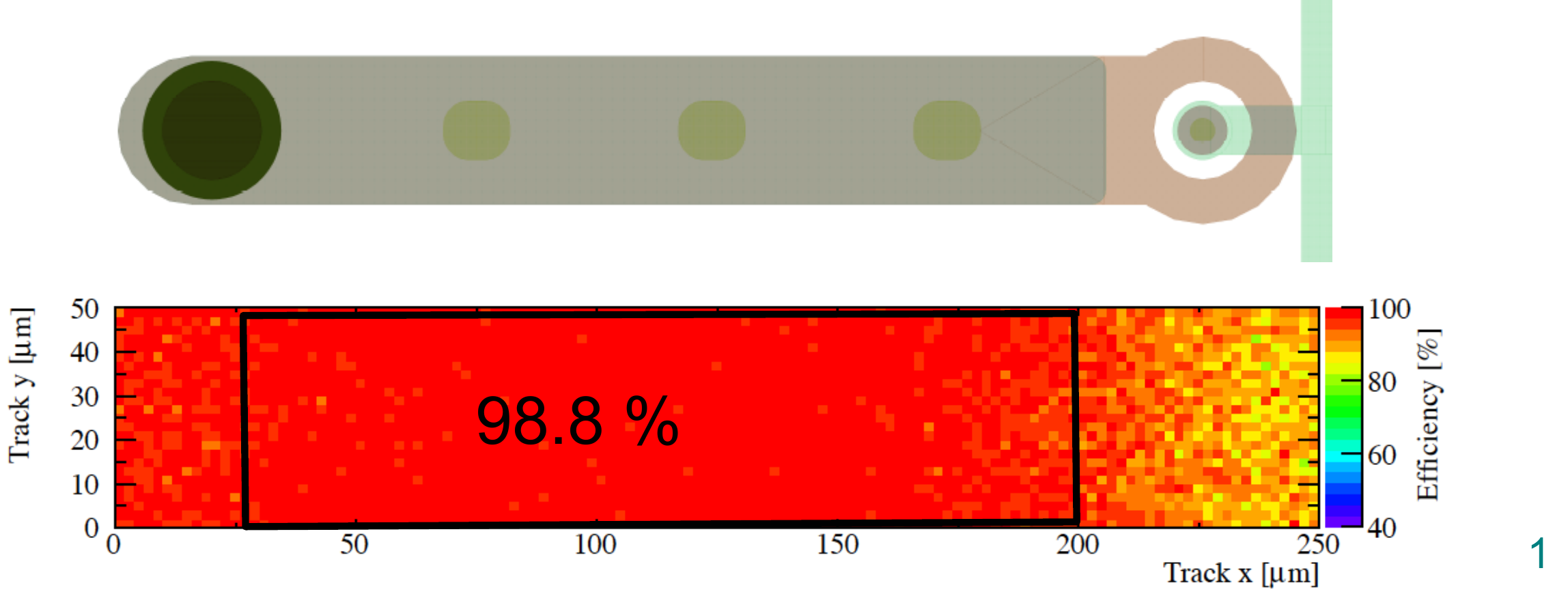}
\caption{Hit efficiency map of the FE-I4 pixel module with a 100 $\mu$m thick sensor, irradiated at a fluence of 
 $\Phi$=$2\times 10^{15}$ \neqcm, obtained in a beam test with 4 GeV electrons at DESY. }
\label{fig:MCZ_PT}
\end{figure}

 Fig.\ref{fig:MCZ_PT} shows the values of the hit efficiency of the 100 $\mu$m thick sensor for all tracks projected into one single pixel cell, 
as a function of the impact point predicted by the beam telescope. The lower global efficiency of this module was traced back to the pixel area 
corresponding to the punch-through structure
and bias rail.
The contours of the low efficiency region are smeared due to the effect of the 
not optimal telescope pointing resolution for this particular setup, estimated to be 27 $\mu$m. This value has been obtained 
from the residual distribution of single hit clusters after disentangling the contribution
of the pixel intrinsic digital resolution.
The module with the 200 $\mu$m thick sensor, measured in the same conditions, does not show instead a decrease of the hit efficiency
in the area of the punch-through structure and bias rail, resulting in an overall higher hit efficiency.
The partial loss of efficiency in the pixel biasing structure has been observed also for other thin pixel production and for pixel sensors of standard thickness at higher fluences \cite{IWORID, thesis_philipp, IBLmodule}. 
Parallel activities on the production of active edge planar pixel sensors are pursued also by other groups, for example
with the fabrication at FBK of FE-I4 compatible sensors with a similar approach to the one followed at VTT, with the 
main difference being in the technology adopted for the trench doping \cite{bomben}.

\section{Vertical Integration Technologies}

\begin{figure}[h] 
\centering 
\includegraphics[width=\columnwidth,keepaspectratio]{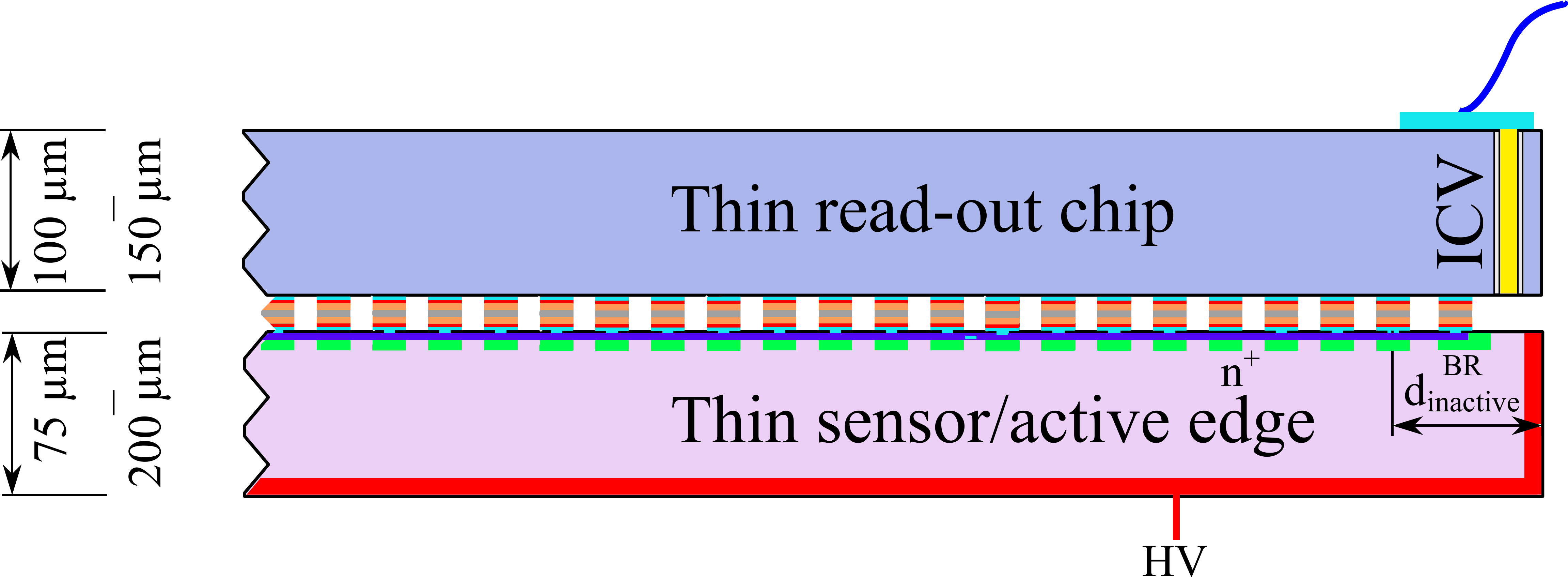}
\caption{Sketch of an hybrid pixel module that could be tiled along all four sides, composed by thin sensors
with active edges and read-out chips where the signals are routed from the front- to the back-side 
by Inter Chip Vias.}
\label{fig:VI_concept}
\end{figure}

Active edge pixels represent a successfull technology to reduce the dead area of tracking detectors but to
obtain modules that can be tiled along all four sides, also the read-out chips have to be redesigned to 
eliminate or strongly diminish the inactive regions. Vertical integration technologies can be used to solve these
problems, applying Inter Chip Vias (ICV) to transport the signals from the front- to the back-side. In this way 
hybrid pixel modules could be built with chips without  the cantilever where the wirebond pads are normally located.
Fig\ref{fig:VI_concept} shows the concept of an hybrid pixel modules, assembled with active edge pixels and the read-out chip
with ICVs.

\begin{figure}[h] 
\centering 
\includegraphics[width=\columnwidth,keepaspectratio]{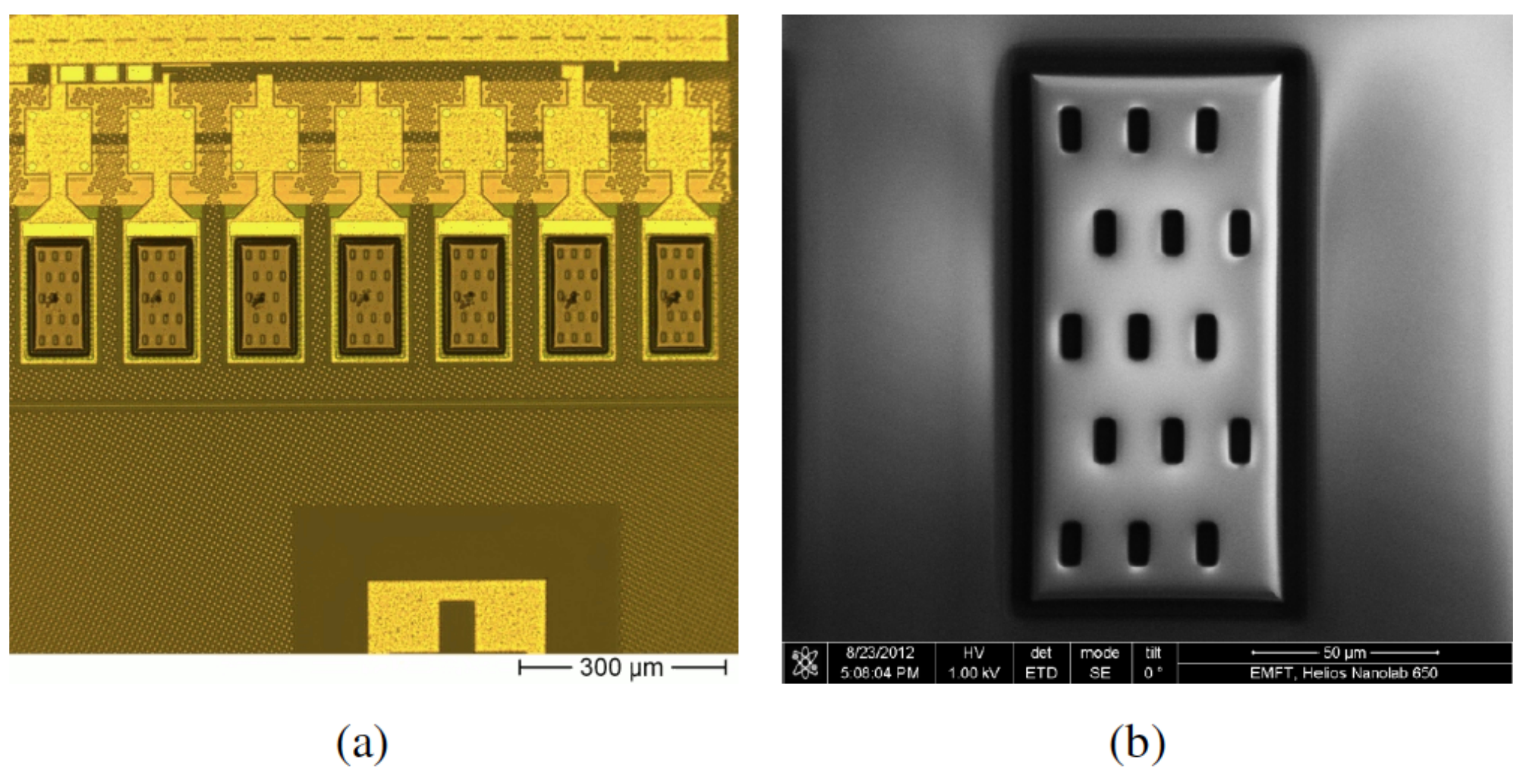}
\caption{(a) Distribution of the ICV over the wire-bonding pad and layout of
the encircling trenches. (b) View on the back-side of the FE-I3 chip after thinning the chip to a thickness of 
60 $\mu$m  in the area corresponding to one wire bonding
pad in the front-side. }
\label{fig:ICV}
\end{figure}

A project about the application of vertical integration technologies to tracking detectors was initiated to demonstrate the use of ICVs together with the Solid Liquid Interdiffusion (SLID) interconnection,
 studied as a possible alternative to bump-bonding.  These two technologies have been developed by the Fraunhofer Institute EMFT and
the processing described in the following has been performed in their clean-room facilities.
SLID is based on the formation of  Cu-Sn alloys where the interconnection takes place at temperatures significantly lower than the ones that
the alloy  can withstand \cite{klumpp}. Therefore it allows for stacking of different interconnected chip and sensor layers without destroying the pre-existing bonds.
The SLID bond pads can have an arbitrary shape and size, with the only constraint that its dimensions exceed 5 $\mu$m by 5  $\mu$m.
In previous runs, a successfull SLID interconnection of FE-I3 chips to pixel sensors with 75 $\mu$m thickness has been demonstrated in the chip
to wafer approach \cite{Pixel2012}. 
In a successive step of this  R\&D program, FE-I3 chips  are dedicated to the assembly of a demonstrator module for ICVs, still using the SLID interconnection
to pixel sensors with  75 $\mu$m and 150 $\mu$m active thickness. 
Since this read-out chip was not designed for the usage with ICVs, these have been etched at the location of the original wire-bonding pads
where the volume below is not filled with logic blocks. Several redundant ICVs were etched into each wire-bonding pad as
illustrated in Fig.\ref{fig:ICV}(a). To insulate the volumes of neighbouring wire-bonding pads against each other, an additional encircling trench is etched.
An optimization of the  process on test wafers has resulted in choosing the cross-section of the ICVs to be 3$\times$10 $\mu$m$^2$.
The ICVs have been first etched using DRIE with columns reaching a depth of 60 $\mu$m. Subsequentely they are isolated 
with Chemical Vapor Deposition (CVD) of silicon dioxide and filled with tungsten CVD to  create the conductive pin between front and back-side
\cite{klumpp}.
The FE-I3 chip wafer has been then thinned to 60 $\mu$m of thickness after having attached a support wafer on the front-side,
in such a way that the ICVs are exposed, as shown in Fig.\ref{fig:ICV}(b).
 
\begin{figure}[h] 
\centering 
\includegraphics[width=\columnwidth,keepaspectratio]{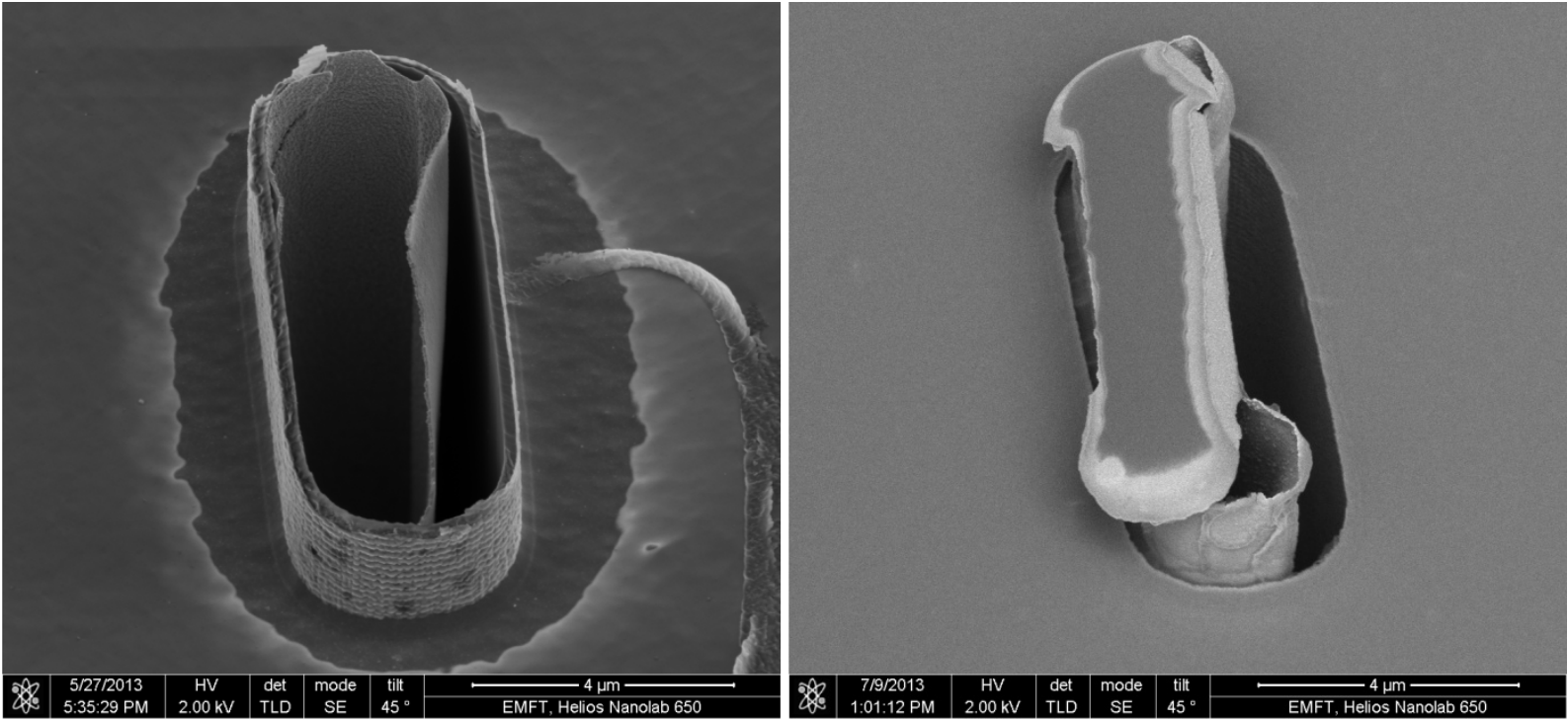}
\caption{SEM images of the ICV bottom as seen on the wafer backside, after recess of the CVD silicon dioxide. The ICVs are hollow
and only a 60 nm layer of TiN around the ICV is visible.}
\label{fig:Hollow}
\end{figure}

A removal of the back-side isolation dioxide has finally been performed, with the recess of the CVD silicon dioxide 
at the bottom of the ICV to create the contact to the underlying tungsten. 
At this stage it has been observed that the tungsten deposition has not been successfull and the ICVs are hollow. 
In Fig.\ref{fig:Hollow}  the ICVs after recess of the CVD silicon dioxide are shown. The walls of the ICVs are coated only
with a 60 nm thick layer of TiN, deposited before the tungsten as an adhesion layer. The filling of ICVs with tungsten
has been previously demonstrated by EMFT with this Vias cross-section in other productions and with test-wafers.
The aging of the tungsten gas used during the Chemical Vapor Deposition has been identified as a possible cause of this
failure.

Still the interconnection of these chips to the FE-I3 compatible sensors with SLID is planned to 
investigate the feasibility of handling such thin structures without affecting their mechanical and electrical
properties.  
The wire bonding of the modules is still possible
thanks to a fan-out structure foreseen as back-up solution on the sensor side.
A second project to create ICVs on the FE-I4 chips has been recently started. For this R\&D activity a new ICV process has to be developed,
because the etching needs to be performed starting from the back-side, contrary to the FE-I3 case, given the presence
of many metal layers below the wire bonding pads. The ICVs will contact the first metal layer of the chip, internally connected
in the fan-out structure to the last metal on top of the wire bonding pad.  The target final thickness of the FE-I4 chips
is in the range of (100-150) $\mu$m.

\section{Conclusions}
Active edge planar pixel sensors, 100 and 200 $\mu$m thick, interconnected to FE-I3 and FE-I4 chips, have been characterized 
before and after irradiation up to a fluence of $5 \times 10^{15}$ \neqcm. Good charge collection properties
and hit efficiencies in beam tests with 4 GeV electrons at DESY have been obtained. Given their narrow inactive region and their
reduced thickness, they represent good candidates to instrument the inner layers of the upgraded ATLAS pixel detector at 
HL-LHC.
An R\&D activity to obtain Inter Chip Vias (ICVs) on the ATLAS read-out chip has started in collaboration with the Fraunhofer Institute EMFT.  This step is meant to prove the feasibility of the signal transport to the newly created readout pads on the backside of the chips allowing for four side buttable devices without the presently used cantilever for wire bonding.  The ICVs are planned to be used in combination with the SLID interconnection in a new run to build a demonstrator module assembled with
FE-I4 chips and thin n-in-p planar pixel sensors.
\section{Acknowledgements}
\label{sec:acknowledgment}
This work has been partially performed in the framework of the CERN RD50 Collaboration. 
The authors thank  V.~Cindro  for the irradiation at JSI,
A. Dierlamm for the irradiation at KIT 
and S. Seidel (University of New Mexico) for the irradiations at LANSCE.
The irradiations at KIT were supported by
the Initiative and Networking Fund of the Helmholtz Association, contract HA-101 (”Physics at the Terascale”).
The irradiation at  JSI and the beam-tests have received funding from the 
European Commission under the FP7 Research Infrastructures project AIDA, grant agreement no. 262025. 
Beam test measurements were conducted within the PPS beam test group comprised by: 
M.~Backhaus,
M. ~Benoit,
M.~Bomben,
G. ~Calderini,
K.~Dette, 
M.~Ellenburg, 
D.~Forshaw,
Ch.~Gallrapp,
M.~George,
S. ~Gibson,
S. ~Grinstein,
J.~Idarraga, 
J.~Janssen,
Z. ~Janoska,
J.~Jentzsch,
O. ~Jinnouchi,
R.~Klingenberg, 
T. ~Kishida,
A.~Kravchenko, 
T.~Kubota, 
T.~Lapsien, 
A.~La Rosa, 
V. ~Libov,
A.~Macchiolo,
G.~Marchiori, 
D. ~Muenstermann,
R.~Nagai,
C.~Nellist, 
G. ~Piacquadio,
R.~Plumer, 
B.~Rastic, 
I.~Rubinskiy,
A.~Rummler,
Y.~Takubo, 
S.~Terzo, 
K.~Toms, 
G.~Troska, 
S. ~Tsiskaridtze,
I. ~Tsurin, 
Y.~Unno,
R.~Wang, 
P.~Weigell,
J.~Weingarten and
T. ~Wittig.


\bibliographystyle{model1-num-names}
\bibliography{<your-bib-database>}

\end{document}